# Configurable Runtime Orchestration for Dynamic Data Retrieval in Distributed Systems

**Abhiram Kandiraju**
Distinguished Engineer, Capital One
United States

## Abstract

Modern enterprise platforms increasingly depend on distributed microservices, analytical data platforms, and external APIs to construct composite responses for applications and operators. Orchestrating data retrieval across these heterogeneous systems remains difficult because many widely used workflow platforms rely on predefined workflows, explicit state-machine definitions, or code-level orchestration models. Apache Airflow centers on DAG-based workflow execution [1], AWS Step Functions defines workflows as state machines using Amazon States Language [2], and Temporal provides durable workflow execution through code-defined workflows [3].

This paper presents a configuration-driven runtime orchestration framework for dynamic data retrieval in distributed systems. The proposed framework generates execution graphs dynamically from configuration at request time, enabling low-latency, per-request orchestration without redeploying workflow code when service integrations evolve. The execution planner performs dependency-aware scheduling and parallel execution of independent tasks, allowing orchestration latency to scale efficiently across distributed service integrations. Unlike traditional orchestration systems optimized for scheduled pipelines, durable business processes, or predefined workflow definitions, the proposed approach focuses on request-driven aggregation across heterogeneous services, including REST APIs, internal microservices, and analytical data platforms. REST-style architectures are particularly relevant in this context because they emphasize scalability, independent evolution of components, and efficient networked interactions [4].

The paper describes the architecture, execution model, and operational tradeoffs of this framework, and presents a representative enterprise case study for **Customer 360** retrieval. It further compares the proposed model with Airflow, Step Functions, and Temporal, arguing that configuration-driven runtime graph generation is especially useful for fast-changing enterprise integrations where flexibility and latency both matter.

## 1. Introduction

Enterprise applications increasingly operate as compositions of microservices, third-party APIs, internal platforms, and analytical data systems, reflecting broader architectural trends in modern data-intensive systems [7]. In many domains, including servicing, operations, fraud detection, compliance, and case management, the useful response is not produced by one service alone. Instead, it is assembled from multiple upstream systems with different protocols, latency profiles, ownership models, and change cycles. Microservices architectures encourage independently deployable services that interact through well-defined APIs, increasing system flexibility but also increasing the need for orchestration across services [8].

This creates a recurring orchestration problem. A single request may need to retrieve account metadata from one service, transactional history from another, fraud or risk indicators from a separate platform, and derived intelligence from an analytical warehouse. In practice, these retrieval patterns evolve often. New services are introduced, dependencies change, optional data sources become mandatory, and conditional logic grows over time. The orchestration layer must therefore balance four goals: flexibility, performance, correctness, and operational maintainability.

Existing workflow platforms such as Apache Airflow [1], AWS Step Functions [2], and Temporal [3] are powerful in adjacent problem spaces, but they are not always ideal for low-latency request-level dynamic retrieval. Airflow models workflows as DAGs with tasks and dependencies, primarily around workflow execution and scheduling. Step Functions models workflows as state machines defined in Amazon States Language. Temporal provides durable execution for reliable long-running workflows through code-defined workflow logic. These are powerful systems, but each assumes that the workflow model itself is defined ahead of execution.

The central claim of this paper is that some enterprise retrieval problems are better addressed by a different model: runtime-generated execution graphs derived from configuration only. In this model, the orchestration engine does not load a precompiled workflow definition for each retrieval use case. Instead, it interprets a configuration specification at request time, builds a dependency graph, executes independent nodes in parallel, and aggregates the results into a final response.

This paper makes the following contributions:

• We introduce a configuration-driven orchestration architecture that generates execution graphs dynamically at runtime rather than relying on predefined workflow definitions.

- We present an execution model that enables low-latency request-level orchestration across heterogeneous APIs, microservices, and analytical data platforms.

- We demonstrate how configuration-driven orchestration allows integration changes to be introduced without redeploying orchestration logic.

- We provide a comparison with existing orchestration systems including Apache Airflow, AWS Step Functions, and Temporal, identifying scenarios where runtime-generated orchestration graphs provide operational advantages.

## 2. Background and Motivation

Distributed enterprise systems commonly integrate over HTTP-based APIs, service contracts, warehouse queries, and derived data products. REST has been especially influential because it promotes loose coupling, scalability, and dynamic substitutability of components in network-based systems[4]. RESTful services have increasingly replaced traditional SOAP-based service models in distributed enterprise systems [5]. In practice, this architectural style helps large organizations evolve systems independently, but it also shifts complexity into orchestration and composition layers.

Three practical forces motivate the need for configuration-driven orchestration.

First, integration volatility is high. Product teams regularly add or revise services, change optionality rules, or alter which data sources participate in a response. Statically encoded orchestration logic turns these changes into deployment events.

Second, request-time composition is increasingly common. Many enterprise experiences require on-demand aggregation rather than overnight or scheduled pipelines. A servicing application, for example, may need to assemble a customer view at interaction time, not hours later.

Third, parallel retrieval matters. When multiple upstream calls are independent, sequential invocation wastes latency budget. Runtime planning can expose parallelism directly from dependency structure.

These needs are related to, but not identical with, the goals of batch scheduling systems or durable long-running workflow engines. The orchestration model discussed here is aimed at fast-moving integration topologies where response composition changes often and where configuration should drive behavior more directly than code deployment.

## 3. Related Work

Workflow orchestration has been widely studied in both academic and industrial systems. Platforms such as Apache Airflow [1] model workflows as Directed Acyclic Graphs representing tasks and dependencies in data pipelines. Cloud-native orchestration services such as AWS Step Functions [2] provide state-machine-based coordination of distributed services. Temporal

[3] focuses on durable execution of long-running workflows using code-defined orchestration models.

These systems provide strong guarantees for workflow scheduling, durability, and operational reliability. However, they generally assume that workflow structures are defined ahead of execution. In contrast, the framework proposed in this paper generates execution graphs dynamically from configuration specifications at runtime. This design enables orchestration logic to evolve through configuration updates rather than workflow redeployment.

## 4. Limitations of Existing Approaches

Apache Airflow is widely used for workflow authoring, scheduling, and monitoring in distributed data pipelines [1]. Its documentation describes the DAG as the central abstraction for representing tasks, dependencies, schedules, and operational behavior. This is highly effective for pipeline-oriented workloads, but less natural for per-request orchestration where the graph itself may differ at runtime.

AWS Step Functions provides state-machine-based orchestration for distributed cloud applications [2] in which workflows are defined using Amazon States Language and executed as event-driven steps. This model supports branching and parallel states, but the workflow structure is still defined in advance. When integration topology changes often, updating the orchestration model remains a workflow-definition task.

Temporal focuses on durable execution and reliable workflow progress for long-running workflows [3]. Its documentation emphasizes crash-proof, scalable workflow execution where code resumes correctly after failure. This is excellent for business processes that must survive long durations and faults. However, it is a different optimization target from request-level orchestration built from runtime configuration.

The issue is not that these platforms are weak. It is that they optimize for different workflow characteristics. For dynamic retrieval orchestration, the main requirement is not merely reliable execution, but **the ability to alter orchestration topology by changing configuration only**, while retaining low operational overhead and exploiting request-time parallelism.

## 5. Proposed Architecture

The proposed framework consists of five logical layers:

1. **Request Adapter**
   Accepts the inbound request and maps user intent or API inputs into orchestration context.

2. **Configuration Resolver**
   Loads the applicable orchestration specification based on operation type, tenant, feature flags, policy constraints, or request metadata.

3. **Execution Planner**
   Parses configuration and constructs a runtime dependency graph containing executable nodes and edges.

4. **Execution Engine**
   Schedules and executes graph nodes, running dependency-free nodes in parallel and honoring ordering constraints where required.

5. **Aggregation and Response Layer**
   Merges outputs, applies transformation rules, validates completeness, and returns a unified response.

## 5.1 Architecture Diagram

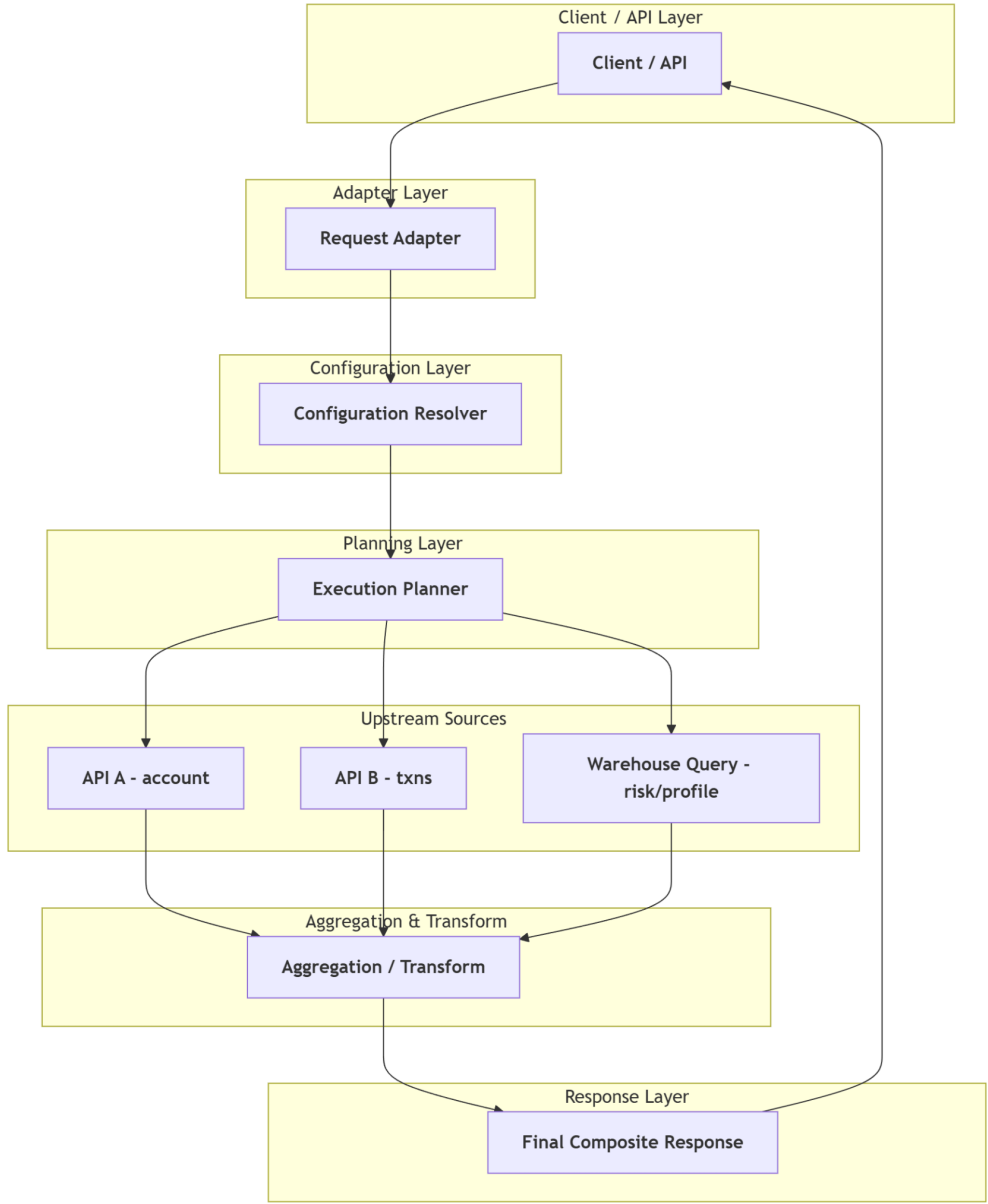

### 5.2 Design Principle

The key design choice is that the execution graph is not stored as a pre-authored workflow artifact. Instead, configuration is treated as the primary source of truth. The planner derives the graph at runtime from declarative step definitions, dependency rules, and execution metadata. This makes the framework closer to a runtime composition engine than a classic workflow compiler.

# 6. Configuration Model

A minimal orchestration specification contains step metadata, operation type, invocation target, input mapping, dependency constraints, timeout budgets, retry policy, and aggregation behavior.

A representative configuration is shown below.

```yaml
operation: customer360
steps:
  - name: accountService
    type: api
    method: GET
    endpoint: /accounts/{customer_id}
    timeout_ms: 500

  - name: transactionService
    type: api
    method: GET
    endpoint: /transactions/{customer_id}
    timeout_ms: 800

  - name: fraudSignals
    type: api
    method: GET
    endpoint: /fraud/signals/{customer_id}
    timeout_ms: 400

  - name: riskProfile
    type: warehouse
    query: SELECT score, segment FROM risk_profiles WHERE customer_id = ?
    timeout_ms: 1200

aggregation:
  strategy: merge
  required_steps: [accountService, transactionService]
```

This model is intentionally declarative. The orchestrator can add a new data source, such as credit exposure or dispute history, by modifying configuration rather than publishing a new workflow implementation.

## 7. Execution Model

At runtime, the planner converts configuration into a dependency graph. Nodes represent invocations or transformations. Edges represent prerequisites. Nodes with zero unresolved dependencies are eligible for immediate execution. The engine maintains a ready queue and executes eligible nodes concurrently up to configured limits.

This planning process resembles DAG-based scheduling approaches used in systems such as Apache Airflow [1], although the graph is generated dynamically rather than defined statically, but differs in one critical aspect: the graph is constructed at request time rather than pre-authored as the workflow artifact. Airflow and similar systems center the DAG as the workflow definition itself. In contrast, this framework derives the graph from configuration after the request arrives.

### 7.1 Execution Pseudocode

```
function execute_operation(request, operation_name):
    config = resolve_configuration(operation_name, request.context)
    graph  = build_execution_graph(config, request)

    ready_queue = nodes_with_no_unmet_dependencies(graph)
    results = {}
    failures = {}

    while ready_queue is not empty:
        parallel_batch = select_executable_nodes(ready_queue)

        batch_outputs = run_in_parallel(parallel_batch, request, results)

        for each output in batch_outputs:
            if output.success:
                results[output.node_name] = output.payload
                mark_node_completed(graph, output.node_name)
            else:
                failures[output.node_name] = output.error
                apply_failure_policy(graph, output.node_name, output.error)

        update_ready_queue(graph, ready_queue, results, failures)

    final_response = aggregate_results(config.aggregation, results, failures)

    validate_response(final_response, config)
    return final_response
```

### 7.2 Failure Semantics

Because retrieval orchestration often spans systems with different reliability characteristics, failure policy must be explicit. The framework distinguishes between:

- required nodes, whose failure blocks a successful composite response
- optional nodes, whose absence may be tolerated with partial results
- fallback nodes, which provide alternate retrieval paths.

This differs from durable execution systems, where the goal is often guaranteed completion of the business workflow. Temporal, for example, is designed around reliable workflow continuation and state persistence across failures [3]. In the proposed model, the primary concern is producing a bounded, request-level result under latency constraints, potentially with partial but valid output.

## 8. Case Study: Customer 360 Retrieval

A representative enterprise use case is **Customer 360 retrieval**, in which a servicing interface needs a unified customer view assembled from multiple upstream systems. A typical request may require:

- account metadata
- transaction history
- fraud indicators
- risk segmentation
- optional recent-case context.

In a traditional orchestration model, adding "recent-case context" would often involve modifying orchestration code or revising a predefined workflow definition. In the proposed framework, the change is expressed by adding a configuration entry and, if needed, a dependency or merge rule. The planner incorporates the new node into the runtime graph automatically.

A concrete flow may proceed as follows:

1. The request adapter receives `customer_id`.
2. The configuration resolver selects the `customer360` orchestration profile.
3. The execution planner builds a graph with four immediately executable nodes: account, transactions, fraud, and risk.
4. The engine executes those nodes in parallel.
5. The aggregation layer normalizes payloads and returns a single composite object.
6. If a fifth optional source is later introduced, the orchestration behavior changes through configuration alone.

This pattern is valuable in enterprise settings where service topology changes often, tenant-specific rules exist, and teams want to reduce redeployment friction.

## 9. Performance Discussion

The framework is intended for environments where flexibility must not come at excessive latency cost. Many analytical data systems rely on well-established principles of concurrency control and transactional consistency in distributed databases [6]. Runtime graph generation introduces some planning overhead, but the overhead is usually small relative to network-bound retrieval operations. The practical performance advantage comes from two sources.

The first is dependency-aware parallelism. Independent calls do not wait for one another simply because they were declared in a sequence. The planner exposes concurrency directly from graph structure.

The second is deployment decoupling. Changes in integration structure do not necessarily require orchestration redeployment, which reduces operational delay and lowers the cost of iterative optimization.

## 10. Comparison with Existing Orchestration Systems

The goal of this comparison is not to rank platforms universally, but to clarify which problem class each system serves best.

| Dimension | Apache Airflow | AWS Step Functions | Temporal | Proposed Framework |
|---|---|---|---|---|
| **Primary model** | DAG-based workflow execution | State-machine workflow orchestration | Durable code-defined workflows | Runtime-generated execution graph |
| **Workflow definition** | Predefined DAG | Predefined state machine | Predefined workflow code | Derived from configuration at request time |
| **Best fit** | Scheduled pipelines, batch workflows | Event-driven workflows, service orchestration | Reliable long-running workflows | Low-latency dynamic data retrieval |
| **Change mechanism** | DAG update/deploy | State machine update | Code change/deploy | Configuration update |
| **Parallelism support** | Yes | Yes | Yes | Yes |
| **Durable long-running execution** | Limited relative emphasis | Moderate | Strong | Not primary goal |
| **Request-level topology changes without redeploy** | Weak | Limited | Limited | Strong |

Airflow explicitly centers the DAG as the model that encapsulates workflow structure and task dependencies [1]. Step Functions centers orchestration around state machines defined using Amazon States Language [2] and Amazon States Language as the workflow definition model. Temporal explicitly centers durable workflow execution defined in application code[3]. The proposed framework differs by making configuration, not workflow code, the principal artifact of orchestration behavior.

## 11. Discussion

This approach has tradeoffs. It is highly flexible, but it shifts responsibility into configuration governance, schema discipline, and observability. Poorly controlled configuration can become difficult to reason about. For that reason, the framework benefits from strong validation, versioning, execution tracing, and policy checks before rollout.

It is also not intended to replace every workflow system. Long-running business processes, retries across days, and strongly durable human-in-the-loop workflows are still better handled by engines specifically designed for durable execution. The contribution here is narrower and more practical: a runtime orchestration model for dynamic enterprise retrieval where integration topology changes frequently and low-latency composition is the main operational requirement.

## 12. Conclusion

Distributed enterprise platforms increasingly require flexible retrieval across APIs, services, and analytical systems. Many established orchestration platforms model workflows as predefined DAGs, state machines, or code-defined durable processes, including Apache Airflow [1], AWS Step Functions [2], and Temporal [3]. Those approaches are valuable, but they do not always align with request-level orchestration in rapidly changing integration environments.

This paper presented a configuration-driven runtime orchestration framework that generates execution graphs dynamically from configuration alone. The approach allows new integrations, dependency changes, and optional retrieval paths to be introduced without redeploying orchestration logic. By combining runtime graph generation, dependency-aware parallelism, and declarative integration specifications, the framework offers a practical architectural model for enterprise systems that require both agility and performance.

**Disclaimer**